\documentclass[aps,prb,twocolumn,letterpaper,superscriptaddress,showpacs]{revtex4-2}

\usepackage{graphicx}
\usepackage{verbatim}
\usepackage{physics}
\usepackage{lineno}
\usepackage{bm}
\usepackage{anyfontsize}
\usepackage[usenames,dvipsnames,x11names]{xcolor}

\usepackage{xr-hyper}
\usepackage[colorlinks=true,linkcolor=blue,filecolor=blue,citecolor=blue,urlcolor=blue]{hyperref}
\newcommand{\figref}[1]{Fig.\,\ref{#1}}

\newcommand{\eqnref}[1]{Eq.\,\eqref{#1}}
\newcommand{\appref}[1]{Appendix~\ref{#1}}

\renewcommand{\Re}{\mathop{\mathrm{Re}}}

\newcommand{\sbf}[1]{{\boldsymbol{#1}}}

\newcommand*{\ab}{\alpha\beta}
\newcommand*{\mn}{\mu\nu}

\newcommand*{\dg}{\dagger}
\newcommand*{\ii}{i}

\makeatletter
\begin{document}
\title{Intrinsic phase fluctuation and superfluid density in doped Mott insulators}
\author{Zeyu Han}
\altaffiliation{These authors contributed equally to this work.}
\affiliation{Institute for Advanced Study, Tsinghua University, Beijing 100084, China}
\author{Zhi-Jian Song}
\altaffiliation{These authors contributed equally to this work.}
\affiliation{Institute for Advanced Study, Tsinghua University, Beijing 100084, China}
\author{Jia-Xin Zhang}
\affiliation{French American Center for Theoretical Science, CNRS, KITP, Santa Barbara, California 93106-4030, USA}
\affiliation{Kavli Institute for Theoretical Physics, University of California, Santa Barbara, California 93106-4030, USA}
\affiliation{Institute for Advanced Study, Tsinghua University, Beijing 100084, China}
\author{Zheng-Yu Weng}
\affiliation{Institute for Advanced Study, Tsinghua University, Beijing 100084, China}

\date{\today}
\begin{abstract}
The doping dependence of the superfluid density $\rho_{\text{s}}$ exhibits distinct behaviors in the underdoping and overdoping regimes of the cuprate, while the superconducting (SC) transition temperature $T_c$ generally scales with $\rho_{\text{s}}$. In this paper, we present a unified understanding of the superconducting transition temperature $T_c$ and $\rho_{\text{s}}$ across the entire doping range by incorporating the underlying mutual Chern-Simons gauge structure that couples the spin and charge degrees of freedom in the doped Mott insulator.  Within this framework, the SC phase fluctuations are deeply intertwined with the spin dynamics, such that thermally excited neutral spins determine $T_c$, while quantum spin excitations effectively reduce the superfluid density at zero temperature. As a result, a Uemura-like scaling of $T_c$ vs. $\rho_{\text{s}}$ in the underdoped regime naturally emerges, while the suppression of both $T_c$ and $\rho_{\text{s}}$ at overdoping is attributed to a drastic reduction of antiferromagnetic spin correlations.
\end{abstract}
\maketitle
\section{Introduction}
One of the key distinctions between the cuprate superconductor and a conventional BCS superconductor lies in the conjecture that the superconducting (SC) phase transition is controlled by superfluid density in the former instead of the Cooper pairing strength in the latter \cite{Emery1995,Carlson2008,Lee2006}. It is experimentally supported by the scaling of $T_c$ with the superfluid density as observed by Uemura \emph {et al. } \cite{Uemura1989,Uemura1992,Uemura2003}. Here strong SC phase fluctuations are generally expected with weak phase stiffness due to small superfluid density, which may be realized at low charge carrier density regime of doped Mott insulators \cite{Anderson1987,Lee2006}. However, a Kosterlitz-Thouless (KT)-type transition temperature with such low superfluid density still predicts a much higher $T_c$ using reasonable parameters from the cuprate \cite{Lee2006,Patrick_normal,Slaveboson_normal_state}, and the issue of why a realistic $T_c$ is so low remains an important puzzle for a high-$T_c$ theory based on the doped Mott insulator description.



The scaling behavior of $T_c$ with the superfluid density persists even into overdoping \cite{Bozovic2016}, where both begin to decrease and eventually vanish with the continuous increase in doping concentration. It challenges a prevailing conjecture that the cuprate evolves into a conventional BCS superconductor with reduced pairing strength in the overdoped regime, as the emergence of a large Fermi surface \cite{Chatterjee,Kyle2005} may indicate a reduction in electronic correlations. In order to explain the reduction in superfluid density based on the BCS theory, some theoretical work has attributed it to strong scattering or disorder effects \cite{Li2021,qianghua_wang,Broun_disorder,PhysRevB.98.054506}. However, in addition to the suppression of $T_c$ and superfluid stiffness, various anomalies still persist, which are non-BCS in nature \cite{Zaanen2016}. The presence of the in-gap states observed by ARPES \cite{10.1038/s41586-021-04251-2} and the gap-filling seen in STM studies \cite{Tromp2023} both indicate that phase fluctuations continue to play a significant role in driving the SC transition. The resonance-like low-energy spin mode is still present in the SC region \cite{ResonancemodeOD,Lipscombe,Capogna2007,Pailhes2006}. At high temperatures, the strange metal behavior has been observed across the entire doping range, from the underdoped to the overdoped side \cite{Ayres2021}. A great deal of experimental evidence seems to point to a unified SC dome bounded by the anomalous phase fluctuations, highlighting the need for a consistent theory. 

In this paper, we discuss an unconventional phase fluctuation that can crucially influence the SC transition temperature $T_c$ and the superfluid density $\rho_{\text{s}}$ in doped Mott insulators. In the conventional slave-boson mean-field theory \cite{Lee2006,Slaveboson_normal_state} of the d-wave SC state in the $t$-$J$ model, the ground state is depicted by a Bose condensate of spinless holons in a short-range AFM ordered or resonating-valence-bond (RVB) spin background, with $T_c$ coinciding with the holon condensation temperature $T_v^0$. However, in the phase-string formalism \cite{Weng_2011} of the $t$-$J$ model, it is found that the holon condensate further suffers from a strong phase frustration from the RVB background if a spin-singlet RVB pair is broken into independent $S=1/2$ spinons. Here, $T_c$ and $\rho_{\text{s}}$ can be substantially renormalized, respectively, by such a phase fluctuation that is excited thermally or quantum mechanically. A generalized Uemura plot is obtained across the entire doping range, which also naturally explains the suppression of both $T_c$ and $\rho_{\text{s}}$ in the overdoped regime as resulting from the destruction of the short-range AFM correlation.

\begin{figure}[h]
\centering
\includegraphics[width=0.95\linewidth]{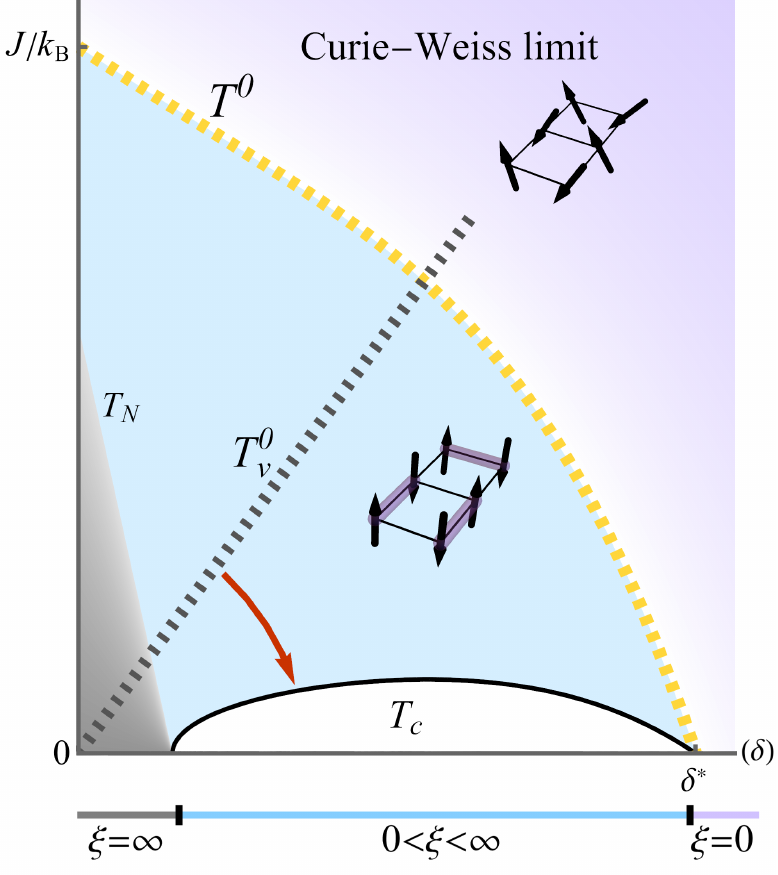}
\caption{The phase diagram for a doped Mott insulator based on the mutual Chern-Simons (MCS) gauge description. The low-temperature white-colored dome indicates the superconducting phase, of which $T_c$ is drastically reduced from a bare $T_v^0$ by the MCS gauge fluctuations. Here the phase coherence is protected by spin local moments forming the bosonic RVB pairing with a finite spin-spin correlation length $\xi$. Self-consistently the antiferromagnetic long-range order (bounded by the Neel temperature $T_N$ with a divergent $\xi$ in the gray area) is quickly destroyed by doped holes and evolves into a short-range ordered phase bounded by $T_0$ (the blue area), which in turn connects to a purple-colored overdoped regime with the Curie-Weiss-like paramagnetic phase of vanishing $\xi$ at doping $\delta^*\simeq 0.26$ \cite{Weng.Ma.2014}.   }           

\label{fig_sketch}
\end{figure}

\begin{figure*}[t]
\centering
\includegraphics*[width=1\linewidth]{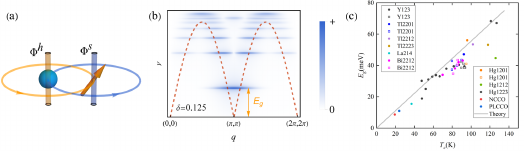}
\caption{(a) Intrinsic phase fluctuations arise from the long-range spin-charge entanglement in the MCS gauge theory: A holon carries a $\pi$-flux $\Phi^h \equiv \pm \pi n^h$ as perceived by a spinon of spin index $\sigma =\pm 1$, and, \emph{vice versa}, a spinon carries a flux $\Phi^s \equiv \pi \sum_\sigma\sigma n^b_\sigma$ seen by holons. (b) A finite density of holons reshapes the $S=1$ spin excitation spectrum into a resonance-like mode at energy $E_g$ near antiferromagnetic wavevector $(\pi, \pi)$ at the mean-field level (with $\delta=0.125$). (c) The superconducting critical temperature $T_c$ according to \eqnref{Tc}, with the experimental data replotted from Ref. \cite{mei_spin-roton_2010}.
}
\label{fig_illustration}
\end{figure*}


\section{Key results}
The phase diagram in \figref{fig_sketch} illustrates the overall evolution of the AFM phase at half-filling upon doping (with doping concentration $\delta$), based on the mutual Chern-Simons (MCS) gauge theory \cite{Weng_2011,Weng.Ma.2014}, in which the SC phase (white-colored dome) is the main focus of the present work. Here the MCS gauge structure refers to an intrinsic sign structure known as the phase string effect in the $t$-$J$ model \cite{Weng.Sheng.1996,Weng1997,Zaanen.Wu.2008}, which arises due to the opening of the Mott gap. This gauge structure implies that the spins (local moments) and doped holes, perceive each other as $\pi$-flux tubes as illustrated in \figref{fig_illustration} (a). Thus, the spin and charge degrees of freedom are intrinsically long-range entangled via an Aharonov-Bohm effect and thereby exert mutual gauge interactions \cite{Weng1997,Zaanen.Wu.2008}. 

At finite doping, the holon condensate will quickly destroy the long-range AFM order so that the low-lying spin-wave excitation around $(\pi,\pi)$ becomes a nonpropagating, resonance-like mode at the Schwinger-boson mean-field level with a gap $E_g$ as shown in \figref{fig_illustration} (b). At finite temperature, thermally excited spins will then introduce strong frustration to the charge condensate via the MCS gauge field depicted in \figref{fig_illustration} (a). In contrast to a bare holon condensation (KT-like) temperature $T_v^0\simeq \rho_{\text{s}}^0\propto \delta t$ (cf. \figref{fig_sketch}), the real SC phase is suppressed as illustrated by the white-colored dome. Specifically, $T_c$ is determined by $E_g$ via a renormalization group analysis as detailed in \appref{RG}, leading to the relation
\begin{equation}\label{Tc}
T_c\simeq \frac{E_g}{6.44k_{\mathrm B}}~,
\end{equation}
which shows excellent agreement with the cuprate superconductors in \figref{fig_illustration} (c), where the resonance energy $E_g$ is either taken from the neutron or Raman scattering measurements \cite{mei_spin-roton_2010}. 

Owing to the same MCS gauge fluctuation, the superfluid density is renormalized by the spin excitations at $T=0$ as follows:
\begin{equation}\label{rho^s}
\rho_{\text{s}} = \rho^0_{\text{s}} \frac{\lambda E_g}{\lambda E_g +\rho^0_{\text{s}}}~,
\end{equation}
where $\lambda$ is a phenomenological parameter. Again, the spin resonance energy $E_g$ plays a crucial role here in the renormalization of the superfluid density. Only in the limit of large $E_g$ can the bare $\rho^0_{\text{s}}$ be restored as the stiffness for the phase fluctuations of the holon condensate. In the opposite limit of $E_g \rightarrow 0 $, $\rho_{\text{s}}\rightarrow \lambda E_g$ similar to $T_c$ in \eqnref{Tc}. Thus, the ratio of $T_c$ to $\rho_{\text{s}}$ is generally expected to be a constant $O(1)$ on the two sides of the SC dome.  The detailed behaviors of $T_c$ and $\rho_{\text{s}}$ vs. doping $\delta$ are shown in \figref{fig:TcNsUemura} (a). A generalized Uemura plot of $T_c$ vs. $\rho_{\text{s}}$ is then obtained in \figref{fig:TcNsUemura} (b).

As summarized in \figref{fig_sketch}, the SC phase composed of the holon condensate and the spin RVB state in the white-colored region is self-consistently protected by an emergent resonance-like spin gap $E_g$ with a finite spin-spin correlation length $\xi$. Above the $T_c$ given in \eqnref{Tc}, the thermally excited (deconfined) spinons in the MCS gauge theory destroy the SC phase coherence, giving rise to strong SC phase fluctuations between $T_c<T<T_v^0$. Here, both $T_c$ and superfluid density vanish as $E_g\rightarrow 0$ with a divergent $\xi$ approaching the AFM ordered phase in underdoping. In contrast, both $T_c$ and the superfluid density vanish near $\delta=\delta^*$ ($T=0$) or $T=T_0$ in overdoping, as shown in \figref{fig:Bozovicdata}. This behavior corresponds to the collapse of the bosonic RVB state with $\xi \rightarrow 0$ and $E_g\rightarrow 0$. Thus, the novel collective fluctuations of the MCS gauge fields, beyond the Schwinger-boson mean-field description, systematically determine the low-temperature phase diagram upon doping.  
\section{Mutual Chern-Simons gauge theory of the doped Mott insulator}
The minimal MCS field-theory description of the mutual $\pi$-flux attachments \cite{Weng.Kou.2003yuc,Kou,Weng.Ye.2011}, based on the phase-string formulation \cite{Weng1997,Weng_2011} of the $t$-$J$ model, is given by the lattice Euclidean Lagrangian $L=L_h+L_s+L_{\text{MCS}}$ as follows:
\begin{eqnarray}
	L_h&=&
	\sum_i h_i^{\dagger}\left[ \partial_0 - i A_0^s(i) - i A_0^e (i)+\mu_h\right] h_i\notag\\
	&\;&-t_h \sum_{i\alpha}\left[h_i^{\dagger} h_{i+\alpha} e^{i {A}_{\alpha}^s(i)+i  {A}_{\alpha}^e(i) }+\text{h.c.}\right],\label{Lh}\\
	L_s&=&
	\sum_{i \sigma} b_{i \sigma}^{\dagger}\left[ \partial_0 - i \sigma A_0^h\left(i\right)+\lambda_b\right]b_{i \sigma}\notag\\
	&\;&-J_s \sum_{i\alpha \sigma}\left[b_{i \sigma}^{\dagger} b_{i+\alpha, \bar \sigma}^{\dagger} e^{i \sigma {A}_{\alpha}^h (i)}+\text {h.c.}\right],\label{Ls}\\
	L_{\mathrm{MCS}}&=& \frac{i}{\pi} \sum_i \epsilon^{\mu \nu \lambda} A_\mu^s(i) \partial_\nu A_\lambda^h (i)\label{LCS},
\end{eqnarray}
in which $L_h$ and $L_s$ describe the dynamics of the matter fields — bosonic spinless holon $h_i$ and bosonic neutral spinon $b_{i\sigma}$ (with $\bar \sigma\equiv-\sigma$), respectively. The indices $\alpha$ and $\beta$ denote the spatial components $x$ and $y$, while $\mu = (\tau, \boldsymbol{r})$ labels the full time-space vector. $\lambda_b$ and $\mu_h$ are the chemical potentials for the spinon $b$ and holon $h$. The hole doping concentration $\delta$ is introduced via $\mu_h$. The renormalized hopping strength $t_h$ and effective spin AFM coupling $J_s$ are determined at a generalized mean-field level in Ref. \cite{Weng.Ma.2014}. The external electromagnetic vector potential ${A}^e$ with a field strength $B^e$ perpendicular to the 2D plane appears in \eqnref{Lh}, and couples to the charge (holon) degree of freedom under the convention of unit charge.

Here the holon field $h$ and spinon field $b_\sigma$ minimally couple to the gauge fields, $A^s_\mu$ and $A^h_\mu$, respectively, which are entangled by the MCS topological term in \eqnref{LCS}. The strengths of these MCS fields are determined by the following equations of motion for $A_0^s$ and $A_0^h$, respectively:
\begin{eqnarray}
      &\epsilon^{\alpha \beta} \partial_\alpha {A}_\beta^h \equiv {B}^h
	= \pi n^h,\label{conAh} \\
	 &\epsilon^{\alpha \beta} \partial_\alpha {A}_\beta^s \equiv {B}^s
	= \pi \sum_\sigma \sigma n_{\sigma}^b .\label{conAs}
\end{eqnarray}
Namely, the holon number $n^h$ and spin density $\sigma n^b_\sigma$ respectively determine the gauge-field strength of $A^h_\mu$ and $A^s_\mu$ as if each matter particle (holon or spinon) is attached to a fictitious $\pi$ flux tube that acts only on the other species as illustrated in  \figref{fig_illustration} (a). 

\subsection{Emergent bosonic RVB as a short-range AFM state}
At half-filling, the action $L$ reduces to $L_s$ in \eqnref{Ls} in the Schwinger-boson mean-field state with $n^h={B}^h=0$. It describes the AFM phase well with $T_0\sim J/k_{\mathrm B}$ denoting the onset temperature of the short-range AFM correlation. Upon doping, a finite flux ${B}^h$ is perceived by the spinons in \eqnref{Ls} according to \eqnref{conAh}, which can drastically modify the spin dynamics as illustrated in \figref{fig_illustration} (b) calculated self-consistently at the \emph{bosonic} RVB mean-field level (cf. Ref. \cite{Weng.Ma.2014}). It shows that the low-lying gapless spin wave excitation (dashed curve) is replaced by a prominent resonance-like mode with gap $E_g$ around the AFM wavevector $(\pi,\pi)$, while the high-energy excitations are dispersed into non-propagating weaker incoherent modes. Namely, an AFM-ordered spin background at half-filling can be turned into a short-range AFM due to the flux-binding effect. A systematic evolution of the spin correlations/dynamics as a function of doping is schematically illustrated in \figref{fig_sketch}, in which the inherited Schwinger-boson-like bosonic RVB order parameter, characterizing the short-range AFM order at $T\leq T_0$, is continuously suppressed with increasing doping. It eventually terminates at some overdoped concentration $\delta^*\simeq 0.26$ at zero temperature or a finite $T_0$, beyond which a Curie-Weiss-like paramagnetic phase sets in \cite{Weng.Ma.2014}. 

\subsection{Superconducting phase coherence}
The condensed charge (holons) would remain SC phase coherent below $T_v^0$, if the internal gauge field $A^s\equiv 0$ in \eqnref{Lh}. Here $T_v^0$ denotes the conventional KT transition temperature for a 2D hard-core holon gas with a bare superfluid density $\rho_{\text{s}}^0$. Without $A^s$, $T_v^0$ would be 
unrealistically high compared to the observed $T_c$ in the slave-boson RVB theory \cite{Patrick_normal,Slaveboson_normal_state}. However, in the present MCS theory, $T_c$ can be drastically reduced by the fluctuations of $A^s$, which is described the $\pi$ flux tubes associated with individual spins in the background according to \eqnref{conAs}. Note that in the RVB ground state, $A^s$ cancels out to ensure phase coherence without generating net fluxes according to \eqnref{conAs}. 

\begin{figure}[t]
	\flushleft
    \includegraphics[width=0.93\linewidth]{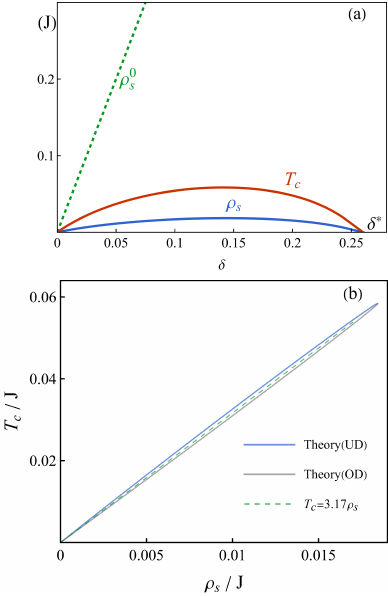}
	\caption{(a) Bare superfluid density $\rho_{\text{s}}^0$, renormalized superfluid density $\rho_s$, and superconducting transition temperature $T_c$ versus doping concentration $\delta$, where $\delta^*\approx 0.26$ denotes the superconducting ending point. The underlying parameters are the same as used at the mean-field level in Ref. \cite{Weng.Ma.2014}.
    (b) $T_c$ versus $\rho_{\text{s}}$, where the blue curve and gray curve represent the underdoped (UD, $\delta<0.14$) and overdoped (OD, $\delta>0.14$) regimes, respectively. The green dashed line indicates the fitting of $T_c=3.17\rho_s$, with $\lambda=1/(2\pi^2)$.}
\label{fig:TcNsUemura}
\end{figure}

On the other hand, once the spinons are thermally excited by breaking up the RVB pairs in \eqnref{Ls}, their associated $\pi$ fluxes in $A^s$ can eventually proliferate through a confinement-deconfinement transition, which results in a non-SC phase full of composite spinon-$\pi$-vortices in the holon condensate described by $L_h$ in \eqnref{Lh}. Such a strong SC fluctuating state is known as the lower pseudogap phase. The corresponding phase transition $T_c$ can be obtained via a renormalization group calculation (cf. Refs. \cite{mei_spin-roton_2010} and \cite{song_thermal_2024}, as well as \appref{RG}), which is essentially determined by the low-lying spin resonance energy $E_g$ as given in \eqnref{Tc}. The universal scaling law of the $T_c$ formula in \eqnref{Tc} is illustrated in \figref{fig_illustration} (c), alongside the experimental data obtained by neutron scattering and Raman scattering measurements \cite{mei_spin-roton_2010}. As noted above, $E_g$ as a function of doping is calculated by a self-consistent mean-field approach, which determines $T_c$ via \eqnref{Tc} as shown in \figref{fig:TcNsUemura}.

\subsection{Superfluid stiffness at zero temperature}
The superfluid density $\rho_{\text{s}}$ characterizes the response of the SC phase to an applied external gauge field $A^e$. According to \eqnref{Lh}, the holon condensate is coupled to both the external $A^e$ and an internal topological gauge field $A^s$. The latter should be integrated out in order to determine the renormalized $\rho_{\text{s}}$. According to the relation \eqnref{conAs}, the flux strength ${B}^s$ for $A^s$ is tied to the spin excitations, implying that the dynamics of $A^s$ is governed by the dynamical spin correlation as follows:
\begin{equation}
D^{A_s}_{\ab}\left(\sbf{q},\ii \omega\right) \equiv \left\langle A^s_\alpha A^s_\beta \right\rangle= \delta_{\ab} \frac{4 \pi^2}{q^2 a^4} \chi^{zz}\left(\sbf{q},\ii \omega\right).
\label{DA}
\end{equation}
Here $\chi^{zz}=\left\langle S^z S^z \right\rangle$ represents the dynamical spin susceptibility calculated using the $b$-spinon Lagrangian $L_s$ in \eqnref{Ls}. At long wavelength, \eqnref{DA} gives a general scaling relation with $E_g$,
\begin{equation}
	\lim_{q\rightarrow0}  \frac{4 \pi^2}{q^2a^2} \chi^{zz}\left(\sbf{q},\ii \omega\right) \propto \frac{1}{\lambda E_g},
\end{equation}
while $\lambda=O(1/2\pi^2)$ is treated phenomenologically, with its order of magnitude detailed in \appref{MF}.

In the SC state, by expressing $h_i=\sqrt{n^h} e^{i \theta^h_i}$ for the holon condensate, the effective Lagrangian can be rewritten in the continuous limit, as follows:
\begin{equation}\label{eqn:LhTotal}
	\begin{split}
		L_{\text{eff}} &=\frac{1}{2}\int_0^\beta\dd\tau\int\dd \boldsymbol{r}\ [\rho_s^0 \left( \partial_\alpha \theta^h - A_\alpha^s - A^{\text{ext}}_\alpha\right)^2+\lambda E_g A^{s2}_\alpha]
	\end{split}
\end{equation}
where we have incorporated the term for the internal gauge field $A^s$ after integrating out the $b$-spinons as discussed above. Subsequently
the equation of motion leads to  $A^s_\alpha = -\frac{\rho_{\text{s}}^0}{\rho_{\text{s}}^0 + \lambda E_g}A_\alpha^{\text{e}}$ ($\rho_{\text{s}}^0\equiv 2t_h n^h$), resulting in
\begin{equation}
\label{eqn:LeffExt}
L_{\text{eff}}=\int\dd \boldsymbol{r}\frac{\rho_{\text{s}}}{2}(A^{\text{e}})^2,
\end{equation}
where $\rho_{\text{s}}$ is given in \eqnref{rho^s} (cf. \appref{effective action} for details).

\begin{figure}[t]
	\includegraphics[width = 0.9\linewidth]{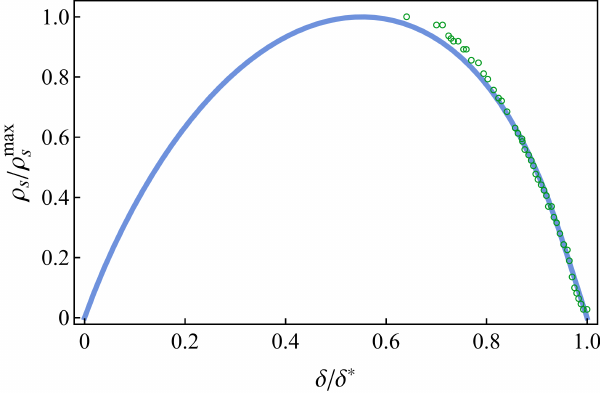}
     \vspace{-0.4cm}
     \caption{Experimentally observed superfluid density (open circle) is suppressed in the overdoped regime, which fits well with the rescaled quantity $\rho_{\text{s}}/\rho_\text{s}^\text{max}$ plotted against the normalized doping level $\delta/\delta^*$, where $\rho_\text{s}^\text{max}$ denotes the maximal $\rho_{\text{s}}$ where $\rho_\text{s}^\text{max}$ denotes the maximal $\rho_{\text{s}}$ shown in \figref{fig:TcNsUemura} (a). The experimental data are taken from Bo{\v z}ovi{\'c} et al. \cite{Bozovic2016}.}
     \vspace{0cm}
	\label{fig:Bozovicdata}
\end{figure}

As shown in \figref{fig:TcNsUemura} (a), $T_c$ and $\rho_{\text{s}}$ are strongly renormalized by the spin resonance energy $E_g$, with $\rho_{\text{s}}$ substantially reduced from the bare value $\rho_{\text{s}}^0$. Especially in the overdoping regime, the doping dependence of $\rho_{\text{s}}$ (cf. \figref{fig:Bozovicdata}) deviates from that of the bare value $\rho_{\text{s}}^0$, decreasing and vanishing at $\delta^*$, which explains the experimental results of Bo{\v z}ovi{\'c} et al. \cite{Bozovic2016}. Note that $\rho_{\text{s}}^0=4\delta J$ is used by taking $t_h=2J$ \cite{Weng.Ma.2014}, which can also be regarded as an independent phenomenological parameter, in addition to $\lambda E_g$ in \eqnref{rho^s}. Finally, in the limit $E_g \rightarrow 0$ at both the underdoped and overdoped ends, corrections from higher energy levels [cf. \figref{fig_illustration} (b)] to both $T_c$ and $\rho_{\text{s}}$ may become important, which lies beyond the scope of the present approach.

\section{Discussion}
In this work, we have shown that the spin excitations are fundamentally entangled with the charge phase fluctuations in a doped Mott insulator through the MCS gauge structure. The latter arises from the phase-string sign structure of the $t$-$J$ model or the Hubbard model, a feature that originates from the opening of the Mott gap. At low doping, although the local spin moments can greatly outnumber the doped holes, the AFM order can be quickly destroyed by the motion of the latter via the MCS field, resulting in a short-range AFM-ordered state characterized by a resonance-like gap $E_g$. In turn, charge coherence is strongly influenced by low-lying spin excitations at $E_g$ via MCS gauge fluctuations acting as “cheap vortices,” which determine the superconducting critical temperature $T_c$ and the zero-temperature superfluid density $\rho_\text{s}$. Here, the relations among $T_c$, $\rho_\text{s}$, and $E_g$, as well as their doping dependence, are consistent with those observed in cuprate superconductors.

Therefore, the phase diagram discussed in this work is emergent, in which both spin resonance-like excitation and SC phase coherence are parts of a self-organized low-temperature phenomenon upon doping the Mott insulator. At a high temperature above $T_0$ or doping beyond $\delta^*$ at $T=0$ (cf. \figref{fig_sketch}), the bosonic RVB state with the mean field coupling $J_s$ melts in \eqnref{Ls} such that one enters the Curie-Weiss paramagnetic phase, where strong scattering of the charge degree of freedom by the MCS gauge fluctuation is expected \cite{chen2024non} to result in a strange metal behavior. Furthermore, in addition to the $b$-spinon in the MCS Lagrangian $L$, an itinerant spinon associated with the backflow of holons will appear at higher energies \cite{Weng_2011,Weng.Ma.2014,Hourglass}. Incorporating the latter can give rise to an hourglass-like dispersion \cite{Hourglass} replacing the spin resonance mode in \figref{fig_illustration} (b) at the RPA level, consistent with inelastic neutron scattering experiments \cite{Keimer.Pailhes.2004, Dogan.Hayden.2004}. However, $T_c$ will still be determined by the resonance mode at $E_g$, which becomes the lower branch of the hourglass excitation \cite{Hourglass}. Finally, the AFM order with $E_g=0$ and $T_c=0$ persisting in the lightly doped regime \cite{Weng.Kou.2003yuc} of \figref{fig_sketch} will be further explored elsewhere.

\begin{acknowledgments}
We acknowledge stimulating discussions with Zhi-Long Wang. The financial support by MOST of China (Grant No.~2021YFA1402101) and NSF of China (Grant No.~12347107) is acknowledged. J.X.Z was funded by the European Research Council (ERC) under the European Union’s Horizon 2020 research and innovation program (Grant Agreement No. 853116, acronym TRANSPORT): J.X.Z was also supported in part by grant NSF PHY-2309135 to the Kavli Institute for Theoretical Physics (KITP).
\end{acknowledgments}

\section*{Data Availability}
The data that support the findings of this article are openly available \cite{han_2025_17829990}.    

\appendix
\onecolumngrid
\section{Derivation of the effective Lagrangian}
\label{effective action}
\subsection{Effective Model for Dynamics of $A^s$}
The Lagrangian of the mutual Chern-Simons gauge theory consists of three parts, $L=L_s+L_h+L_{\text{MCS}}$, where
\begin{equation}
    L_h=\sum_i h_i^{\dagger}\left[ \partial_0 - i A_0^s(i) - i A_0^e (i)+\mu_h\right] h_i
    -t_h \sum_{i,\alpha}\left[h_i^{\dagger}
    h_{i+\alpha} e^{i \boldsymbol{A}_{\alpha}^s(i)+i\boldsymbol{A}_{\alpha}^e(i) }+\text{H.c.}\right],
\end{equation}
\begin{equation}
	L_s=\sum_{i,\sigma} b_{i,\sigma}^{\dagger}\left[ \partial_0-i\sigma A_0^h\left(i\right)+\lambda_b\right]b_{i,\sigma}
    -\frac{J_s}{2}\sum_{i,\alpha,\sigma}\left[b_{i,\sigma}^{\dagger} b_{i+\alpha,\bar\sigma}^{\dagger} e^{i \sigma \boldsymbol{A}_{\alpha}^h (i)}+\text{H.c.}\right],    
\end{equation}
\begin{equation}
L_{\mathrm{MCS}}=\frac{i}{\pi} \sum_i \epsilon^{\mu \nu \lambda} A_\mu^s(i) \partial_\nu A_\lambda^h (i)
    =\frac{i}{\pi} \sum_i A_0^s(i)\boldsymbol{B}^h(i)-\sum_i\frac{i}{\pi}\boldsymbol{A}^s(i)\cdot (\boldsymbol{E}^h(i)\times\hat{z}),
\end{equation}
represent the holon action, spinon action and the mutual Chern-Simons term respectively. Integrating out the spinon matter field $b$ and $b^\dagger$, leads to an effective action $S_{\text{eff}}$
\begin{equation}
\begin{split}
Z&=\int D[h^\dagger,h,b^\dagger,b,\boldsymbol{A}^s\boldsymbol{A}^h]\exp[-\int_0^\beta \dd \tau L]=\int D[h^\dagger,h] D\boldsymbol{A}^sD\boldsymbol{A}^h \exp(-S_{\text{eff}})
\end{split}
\end{equation}
with
\begin{equation}
\begin{split}
    S_{\text{eff}} = \frac{1}{2} \sum_{i\omega_n}\int \dd^2\sbf{q}A^h_\mu(\boldsymbol{q},i\omega_n)\Pi^s_{\mn}(\boldsymbol{q},i\omega_n) A^h_\nu(-\boldsymbol{q},-i\omega_n)+\int_0^\beta \dd \tau L_h+\int_0^\beta \dd \tau L_{\text{MCS}}.
\end{split}
\end{equation}
The information of the $b$-spinon matter field is encoded in the polarization tensors defined as follows:
\begin{equation}
\begin{split}
\Pi^s_{00}(\boldsymbol{q},i\omega_n) &= \sum_{\sigma \sigma^\prime } \sigma \sigma^\prime \left\langle n^b_\sigma(\boldsymbol{q},i\omega_n) n^b_{\sigma^\prime}(-\boldsymbol{q},-i\omega_n) \right\rangle
= 4\left\langle S^z(\boldsymbol{q},i\omega_n) S^z(-\boldsymbol{q},-i\omega_n)\right\rangle = 4 \chi^{zz}(\boldsymbol{q},i\omega_n),\\
\Pi^s_{0\alpha}(\boldsymbol{q},i\omega_n) &= \sum_{\sigma}2\sigma\left\langle n^b_\sigma(\boldsymbol{q},i\omega_n)j^{s,(P)}_{\alpha}(-\boldsymbol{q},-i\omega_n)\right\rangle,\\
\Pi^s_{\alpha\beta}(\boldsymbol{q},i\omega_n)& = \Pi^{s,(P)}_{\alpha\beta}(\boldsymbol{q},i\omega_n) +\Pi^{s,(D)}_{\alpha\beta}(\boldsymbol{q},i\omega_n)
=4\left\langle j^{s,(P)}_{\alpha}(\boldsymbol{q},i\omega_n) j^{s,(P)}_{\beta}(-\boldsymbol{q},-i\omega_n)\right\rangle
+2\left\langle \frac{\partial j^{s,(D)}_{\alpha}(\boldsymbol{q},i\omega_n)}{\partial \delta A_\alpha^h}\right\rangle\delta_{\alpha\beta}.    
\end{split}
\end{equation}
In the above derivation, $S^z(\boldsymbol{q},i\omega)$ is the Fourier transformation of $S_i^z=\frac{1}{2}(b_{i,\uparrow}^{\dagger}b_{i,\uparrow}-b_{i,\downarrow}^{\dagger}b_{i,\downarrow})$. $j^{s}_{i\alpha}= j^{s,(P)}_{i\alpha} + j^{s,(D)}_{i\alpha}$, $j^{s,(P)}_{\alpha}(\boldsymbol{q},i\omega)$ and $j^{s,(D)}_{\alpha}(\boldsymbol{q},i\omega)$ defined as:
\begin{equation}
\begin{split}
j^{s,(P)}_{i\alpha} & = \frac{ \partial L_s[\bar{\boldsymbol{A}}^h+\delta \boldsymbol{A}^h]}{\partial\delta A^h_{i\alpha}}=i\frac{J_s}{2}\sum_\sigma\left(\sigma b_{i+\alpha,\sigma}^\dg b_{i,\bar\sigma}^\dg e^{i\sigma\bar{A}^h_{i\alpha}}-\text{H.c.}\right),\\
j^{s,(D)}_{i\alpha} & =\frac{ \partial^2 L_s}{\partial A^h_{i\alpha} \partial A^h_{i\beta}} \delta A^h_{i\alpha}=-\frac{J_s}{2}\sum_\sigma \left(b_{i+\alpha,\sigma}^\dg b_{i,\bar\sigma}^\dg e^{iv\bar{A}^h_{i\alpha}}+\text{H.c.}\right),
\end{split}
\end{equation}
are the Fourier transforms of the paramagnetic and diamagnetic components of the total spinon current $j^{s}_{i\alpha}$.

When spinons are in the normal state (no spinon condensate, no antiferromagnetic long-range order), the spinon current is conserved due to the $U(1)$ gauge symmetry of $\boldsymbol{A}^h$. We therefore expect a Maxwell term of $\boldsymbol{E}^h$  and $\boldsymbol{B}^h$ in the  effective gauge theory. Conservation of spinon current $\partial_\mu j^s_\mu =0$ gives constraints on polarization tensors $\partial_\mu \Pi^s_{\mu\nu}=0$, which will be detailed in momentum space in the long-wavelength limit below,
\begin{eqnarray}
	\Pi^s_{00} &=& q^2 \Pi_0^s , \\
	\Pi^s_{0\alpha} &=& (i\omega_n) q_\alpha \Pi_0^s + i  \epsilon^{\alpha\beta} q_\beta \Pi^s_1 ,\\
	\Pi^s_{\alpha 0} &=& (i\omega_n) q_\alpha \Pi_0^s - i \epsilon^{\alpha\beta} q_\beta \Pi^s_1 ,\\
	\Pi^s_{\alpha\beta} &=& \delta_{\alpha\beta} (i\omega_n)^2  \Pi_0^s + \epsilon^{\alpha\beta} \omega_n \Pi^s_1 + \left( \delta_{\alpha\beta}q^2 - q_\alpha q_\beta\right) \Pi^s_2 .
\end{eqnarray}
Without polarization induced by an external magnetic field, time reversal symmetry remains intact, so $\Pi^s_1=0$. $S_{\text{eff}}$ then contains only a Maxwell term,
\begin{equation}\label{eqn:EffGaugeAct}
\begin{split}
    S_{\text{eff}}&=\int_0^\beta\dd\tau\int\dd \boldsymbol{r}(\frac{1}{2}\Pi_0^s \boldsymbol{E}^h\cdot\boldsymbol{E}^h
	+ \frac{1}{2} \Pi^s_2 \boldsymbol{B}^h\cdot\boldsymbol{B}^h)+ \int_0^\beta\dd\tau(L_{h}+L_{\text{MCS}})
\end{split}
\end{equation}
One step further, integrating out internal fields $\boldsymbol{E}^h$ and $\boldsymbol{B}^h$, we have
\begin{equation}
\begin{split}
\int D\boldsymbol{E}^hD\boldsymbol{B}^h \exp(-
S_{\text{eff}})=\exp[-\int_0^\beta\dd\tau\int\dd \boldsymbol{r}[\frac{1}{2\pi^2}\frac{1}{\Pi_0^s}\left(A^s_\alpha\right)^2
+ \frac{1}{2\pi^2}\frac{1}{\Pi_2^s}\left(A^s_0\right)^2]+\int_0^\beta\dd\tau L_h].
\end{split}
\end{equation}
We do not consider the compressibility or response to chemical potential here, thus we take the gapped $A^s_0=0$  fixed at saddle point. Recall that $\Pi_0^s = \frac{\Pi^s_{00}}{q^2}$, integrating over $A^h$ and $b$-spinon leads to an additional term of $A^s$ in the action, which reads
\begin{equation}\label{eqn:app:Lseff}
\begin{split}
&\int_0^\beta\dd\tau\int\dd \boldsymbol{r}\frac{1}{2\pi^2}\frac{1}{\Pi_0^s}\left(A^s_\alpha\right)^2=\int_0^\beta\dd\tau\int\dd \boldsymbol{r}\frac{1}{2} \frac{1}{\pi^2} \left(\frac{\Pi^s_{00}}{q^2}\right)^{-1} \left(A^s_\alpha\right)^2=\int_0^\beta\dd\tau\int\dd \boldsymbol{r}\frac{1}{2}  \left(\frac{4 \pi^2 \left\langle S^z S^z \right\rangle }{q^2}\right)^{-1} \left(A^s_\alpha\right)^2\\
&=\int_0^\beta\dd\tau\int\dd \boldsymbol{r}\frac{1}{2}  \left(\frac{ 4 \pi^2 \chi^{zz}(q)}{q^2}\right)^{-1} \left(A^s_\alpha\right)^2=\int_0^\beta\dd\tau\int\dd\boldsymbol{r}\frac{1}{2} \left( D^{A_s}_{\alpha \alpha}\right)^{-1}\left(A^s_\alpha\right)^2.
\end{split}
\end{equation}

This can be understood as the equivalence between single spin excitation and the related flux $\sbf{B}^s = \pi \sum_\sigma \sigma n^b_\sigma = 2\pi S^z$:
\begin{equation}
\begin{split}
&\chi^{zz}(\boldsymbol{q},i\omega_n)\equiv \left\langle S^z(\boldsymbol{q},i\omega_n)S^z(-\boldsymbol{q},-i\omega_n) \right\rangle = \frac{1}{4\pi^2} \left\langle \boldsymbol{B}^s(\boldsymbol{q},i\omega_n)\boldsymbol{B}^s(-\boldsymbol{q},-i\omega_n)\right\rangle\\
=&\frac{1}{4\pi^2} \left( q_x^2 \left\langle A^s_y(\boldsymbol{q},i\omega_n)A^s_y(-\boldsymbol{q},-i\omega_n) \right\rangle+ q_y^2 \left\langle A^s_x(\boldsymbol{q},i\omega_n)A^s_x (-\boldsymbol{q},-i\omega_n)\right\rangle 
-2 q_x q_y \left\langle A^s_x(\boldsymbol{q},i\omega_n)A^s_y(-\boldsymbol{q},-i\omega_n) \right\rangle \right)\\
=&\frac{1}{4\pi^2} q^2 D^{A_s}_{\alpha \alpha},
\end{split}
\end{equation}
where $D^{A_s}_{\alpha \beta}\equiv \left\langle A^s_\alpha A^s_\beta \right\rangle$ denotes the propagator of gauge field $\boldsymbol{A}^s$, and isotropy leads to $D^{A_s}_{\alpha \beta} = \delta_{\ab} D^{A_s}_{\alpha \alpha}$. We see this relation exactly maps to the effective Lagrangian derived by field integration of spinon \eqnref{eqn:app:Lseff}. 
\subsection{Effective Model for External Response: From \eqnref{eqn:LhTotal} to 
\eqnref{eqn:LeffExt}}
In the long wavelength limit, the effective action of holon and $\boldsymbol{A}^s$ reads,
\begin{equation}
\begin{split}
    S&=\frac{1}{2}\int_0^\beta\dd\tau\int\dd \boldsymbol{r}\ \rho_s^0 \left( \partial_\alpha \theta^h - A_\alpha^s - A^{\text{ext}}_\alpha\right)^2+\int_0^\beta\dd\tau\int\dd\boldsymbol{r}\frac{1}{2} \left( D^{A_s}_{\alpha \alpha}\right)^{-1}\left(A^s_\alpha\right)^2\\
    &=\frac{1}{2}\int_0^\beta\dd\tau\int\dd \boldsymbol{r}\ [\rho_s^0 \left( \partial_\alpha \theta^h - A_\alpha^s - A^{\text{ext}}_\alpha\right)^2+\lambda E_g A^{s2}_\alpha],
\end{split}
\end{equation}
\twocolumngrid
\noindent where $\rho_0^s=2t_hn^h=2t_h\delta$. The equilibrium is given by the saddle point equation of internal gauge field $A^s$ after fixing the gauge to eliminate the matter field,
\begin{equation}
	\frac{\delta S}{\delta A^s_\alpha} = 0 \quad \Rightarrow \quad
	A^s_\alpha = -\frac{\rho_0^s}{\rho_0^s + \lambda E_g}A^{\text{ext}}_\alpha.
\end{equation}
Physically, the minus sign implies that the external field $A^{\text{ext}}$ is screened by the internal gauge field $A^{s}$. Put this saddle point equation back to \eqnref{eqn:LhTotal}, we arrive at the effective action of external field:
\begin{equation}
	S_{\text{ext}}[A^{\text{ext}}_\alpha]= \frac{1}{2}\int_0^\beta\dd\tau\int\dd \boldsymbol{r}\ \rho_0^s \frac{\lambda E_g}{ \rho_0^s + \lambda E_g}\left(A^{\text{ext}}_\alpha\right)^2.
\end{equation}
Besides the saddle point method, a similar result can also be obtained by integrating over $\boldsymbol{A}^s$.

\section{Mean-field theory of $b$-spinon}
\label{MF}
The $b$-spinon ground state and excitations are described by the following Hamiltonian,
\begin{equation}
\begin{split}
    H_s &= -J_s\sum_{i,\alpha,\sigma} \left( b_{i+\alpha,\sigma}^{\dagger} b_{i,\bar\sigma}^{\dagger} e^{i \sigma A_{i\alpha}^h} + h.c. \right)\\
    &+ \lambda_b\sum_{\sigma} b_{i,\sigma}^{\dagger} b_{i,\sigma}+\text{const.},
\end{split}
\end{equation}
where $\alpha=\hat{x},\hat{y}$, $J_s=J_{\text{eff}}\Delta^s/2$ and $J_{\text{eff}}$ is the effective super-exchange in \cite{Weng.Ma.2014}. We take the convention of the Fourier transform as
\begin{equation}
b_{i,\sigma}=\frac{1}{\sqrt{N}}\sum_{\boldsymbol{k}}e^{-i\boldsymbol{k}\cdot\boldsymbol{r}_{i}}b_{\boldsymbol{k},\sigma},\ \boldsymbol{k}\in \text{BZ}.
\end{equation}
The Hamiltonian can be diagonalized, with the notification that each plaquette is attached with $\delta\pi$ flux through the gauge field $A_{i\alpha}^h=A_{i+\alpha,i}^h$ generated by the uniformly condensed holons ($\delta=2p/q$, $p$ and $q$ are coprime integers). By choosing the gauge field on each bond as $ (A^h_{ix},A^h_{iy})=(0,Q i_x)$, where $Q=\delta\pi$ and $i_x$ labels the horizontal position of site $i$, the Hamiltonian can be written in momentum space as
\begin{equation}
H_{s}=\sum_{\boldsymbol{k}\in\text{MBZ}}
\begin{pmatrix}
\phi_{\boldsymbol{k},\uparrow}^{\dagger}&\phi_{-\boldsymbol{k},\downarrow}
\end{pmatrix}
\begin{pmatrix}
\lambda_b I&K\\
K&\lambda_b I
\end{pmatrix}
\begin{pmatrix}
\phi_{\boldsymbol{k},\uparrow}\\
\phi_{-\boldsymbol{k},\downarrow}^\dagger\\
\end{pmatrix}
\end{equation}
where
\begin{equation}
\begin{split}
    \phi_{\boldsymbol{k},\uparrow} &= \left( b_{\boldsymbol{k}-Q \hat{x}, \uparrow}, \cdots, b_{\boldsymbol{k}-(q-1)Q \hat{x}, \uparrow}, b_{\boldsymbol{k}-q Q \hat{x}, \uparrow} \right)^T, \\
    \phi_{-\boldsymbol{k},\downarrow} &= \left( b_{-\boldsymbol{k}+Q \hat{x},\downarrow}, \cdots, b_{-\boldsymbol{k}+(q-1)Q \hat{x},\downarrow}, b_{-\boldsymbol{k}+q Q\hat{x},\downarrow} \right),
\end{split}
\end{equation}
$I$ is the identity matrix and
\begin{equation}
\begin{split}
    &K=-\frac{J_{\text{eff}}\Delta^s}{2}\times\\
    &\begin{pmatrix}
        2 \cos(k_x - Q) & e^{-iky} & 0 & e^{iky} \\
        e^{iky} & 2 \cos(k_x - 2Q) & e^{-iky} & 0 \\
        0 & e^{iky} & \cdots & e^{-iky} \\
        e^{-iky} & 0 & e^{iky} & 2 \cos(k_x - qQ)
    \end{pmatrix}.
\end{split}
\end{equation}
$\sum_{\boldsymbol{k}\in\text{MBZ}}$ denotes a summation over the magnetic Brillouin zone (MBZ), which arises due to the reduced translational symmetry imposed by the flux.

\begin{figure}[h]
\includegraphics[width = 0.8\linewidth]{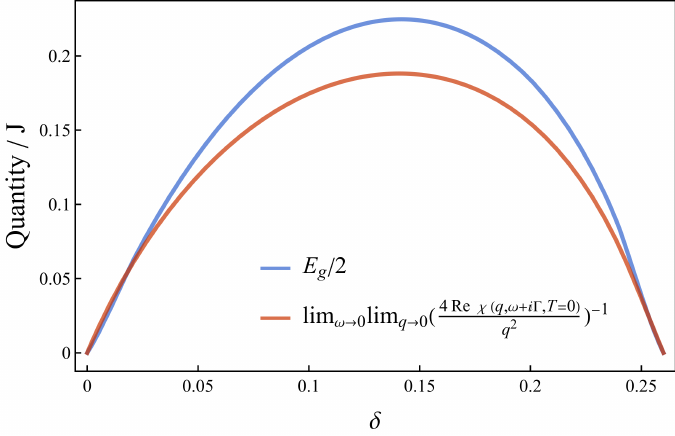}
\vspace{-0.4cm}
\caption{The low momentum, low frequency scaling of $\Re\chi^{zz}$. Numerically, we show that the scaling behavior is $q^2$.}
\vspace{0cm}
\label{chibscaling}
\end{figure}

The $b$-spinon can be transformed first using
\begin{equation}
\begin{split}
 \phi_{\boldsymbol{k},\uparrow}&=\sum_r \omega_{r}(\boldsymbol{k})\beta_{r,\boldsymbol{k},\uparrow},\\
 \phi^{\dagger}_{-\boldsymbol{k},\downarrow}&=\sum_r\omega_{r}(\boldsymbol{k})\beta^{\dagger}_{r,-\boldsymbol{k},\downarrow}.
\end{split}
\end{equation}
$\omega_{r}(\boldsymbol{k})$ is the $r$-th eigenvector of $K$ matrix, satisfying the eigen-equation
\begin{equation}
K\omega_{r}(\boldsymbol{k})=\omega_{r}(\boldsymbol{k})\xi_{r,\boldsymbol{k}}.
\end{equation}
Then the Hamiltonian reads
\begin{equation}
H_{s}=\sum_r\sum_{\boldsymbol{k}\in\text{MBZ}}
\begin{pmatrix}
\beta_{r,\boldsymbol{k},\uparrow}^{\dagger}&\beta_{r,-\boldsymbol{k},\downarrow}
\end{pmatrix}
\begin{pmatrix}
\lambda_b&\xi_{r,\boldsymbol{k}}\\
\xi_{r,\boldsymbol{k}}&\lambda_b
\end{pmatrix}
\begin{pmatrix}
\beta_{r,\boldsymbol{k},\uparrow}\\
\beta_{r,-\boldsymbol{k},\downarrow}^\dagger,\\
\end{pmatrix}.
\end{equation}
The fully diagonalized Hamiltonian reads
\begin{equation}
H_s=\sum_{r,\boldsymbol{k},\sigma}E_{r,\boldsymbol{k}}\gamma^\dagger_{r,\boldsymbol{k},\sigma}\gamma_{r,\boldsymbol{k},\sigma}+\text{const.}
\label{diagonalized b-spinon}
\end{equation}
with
\begin{equation}
\begin{split}
E_{r,\boldsymbol{k}}&=\sqrt{\lambda_b^2-\xi_{r,\boldsymbol{k}}^2},\\
\begin{pmatrix}
\gamma_{r,\boldsymbol{k},\uparrow}\\
\gamma^\dagger_{r,-\boldsymbol{k},\downarrow}
\end{pmatrix}
&=
\begin{pmatrix}
u_{r,\boldsymbol{k}}&v_{r,\boldsymbol{k}}\\
v_{r,\boldsymbol{k}}&u_{r,\boldsymbol{k}}
\end{pmatrix}
\begin{pmatrix}
\beta_{r,\boldsymbol{k},\uparrow}\\
\beta^\dagger_{r,-\boldsymbol{k},\downarrow},
\end{pmatrix}
\end{split}
\end{equation}
where the coherent factors $u_{r,\boldsymbol{k}}$ and $v_{r,\boldsymbol{k}}$ read
\begin{equation}
\begin{split}
    u_{r,\boldsymbol{k}}&=\frac{1}{\sqrt{2}}\sqrt{1+\frac{\lambda_b}{E_{{r,\boldsymbol{k}}}}},\\
	v_{r,\boldsymbol{k}}&=\frac{1}{\sqrt{2}}\textrm{sgn($\xi_{{r,\boldsymbol{k}}}$)}\sqrt{-1+\frac{\lambda_b}{E_{r,\boldsymbol{k}}}}.
\end{split}
\end{equation}
The mean-field self-consistent equations are derived by finding the saddle point of the free energy,
\begin{equation}
\begin{split}
F(\lambda_b,\Delta^s)&= \sum_{r,\boldsymbol{k}} \left[E_{r,\boldsymbol{k}} + \frac{2}{\beta} \ln(1 - e^{-\beta E_{r,\boldsymbol{k}}}) \right]\\
&- 2\lambda_b N + J_{\text{eff}}(\Delta^s)^2 N,
\end{split}
\end{equation}
whose saddle point equations read
\begin{equation}
\begin{split}
\sum_{r,\boldsymbol{k}} \frac{\lambda_b}{E_{r,\boldsymbol{k}}}\coth(\frac{\beta E_{r,\boldsymbol{k}}}{2})&=2N,\\
\sum_{{r,\boldsymbol{k}}}\frac{J_{\text{eff}}\xi_{{r,\boldsymbol{k}}}^2}{4E_{r,\boldsymbol{k}}}\coth(\frac{\beta E_{{r,\boldsymbol{k}}}}{2})&=2N.
\end{split}
\end{equation}
Following the above derivation, the Fourier transformation of $S_i$ reads
\begin{equation}
\begin{split}
S({\boldsymbol{q}})&=\frac{1}{2}\sum_{\boldsymbol{k}}(b^\dagger_{\boldsymbol{k},\uparrow}b_{\boldsymbol{k}+\boldsymbol{q},\uparrow}
-b^\dagger_{\boldsymbol{k},\downarrow}b_{\boldsymbol{k}+\boldsymbol{q},\downarrow})\\
&=\frac{1}{2}\sum_{k\in \text{MBZ}}\sum_{m,n}\omega^\dagger_m(\boldsymbol{k})\omega_n(\boldsymbol{k}+\boldsymbol{q})\times\\
&\begin{pmatrix}
\beta^\dagger_{m,\boldsymbol{k},\uparrow}&\beta_{m,-\boldsymbol{k},\downarrow}
\end{pmatrix}
\begin{pmatrix}
1&0\\
0&-1
\end{pmatrix}
\begin{pmatrix}
\beta_{n,\boldsymbol{k}+\boldsymbol{q},\uparrow}\\
\beta^\dagger_{n,-\boldsymbol{k}-\boldsymbol{q},\downarrow}
\end{pmatrix}
.
\end{split}
\end{equation}
The spin susceptibility is defined in \cite{Chen2005,Weng1999}, which reads
\begin{equation}
 \chi^{zz}(\boldsymbol{q},\tau)=\frac{1}{N\beta}\langle\hat{T} S(\boldsymbol{q},\tau)^zS(-\boldsymbol{q},0)\rangle_{0}^{c}
\end{equation}
where $\langle\hat O\rangle_0$ denotes the expectation value of $\hat{O}$ under the mean-field quadratic Hamiltonian, superscript $c$ denotes the connected contraction. The spin susceptibility in the frequency domain can be obtained,
\begin{equation}
\begin{split}
   &\chi^{zz}(\boldsymbol{q},i\omega_{n})=\int\dd e^{i\omega_n\tau}\chi^{zz}(\boldsymbol{q},\tau)\\
   &=\frac{1}{4N}\sum_{\boldsymbol{k},m,n}
    |\omega_{m}^\dagger(\boldsymbol{k})w_{n}(\boldsymbol{k}+\boldsymbol{q})|^2\times\\
    &[(1-\frac{\lambda^{2}-\xi_{m,\boldsymbol{k}}\xi_{n,\boldsymbol{k}+\boldsymbol{q}}}{E_{m,\boldsymbol{k}}E_{n,\boldsymbol{k}+\boldsymbol{q}}})\frac{E_{m,\boldsymbol{k}}+E_{n,\boldsymbol{k}+\boldsymbol{q}}}{(i\omega_{n})^{2}-(E_{m,\boldsymbol{k}}+E_{n,\boldsymbol{k}+\boldsymbol{q}})^{2}}\\
    &\cdot(1+n_{m,\boldsymbol{k}}+n_{n,\boldsymbol{k}+\boldsymbol{q}})\\
    &+(1+\frac{\lambda^{2}-\xi_{m,\boldsymbol{k}}\xi_{n,\boldsymbol{k}+\boldsymbol{q}}}{E_{m,\boldsymbol{k}}E_{n,\boldsymbol{k}+\boldsymbol{q}}})\frac{-E_{m,\boldsymbol{k}}+E_{n,\boldsymbol{k}+\boldsymbol{q}}}{(i\omega_{n})^{2}-(E_{m,\boldsymbol{k}}-E_{n,\boldsymbol{k}+\boldsymbol{q}})^{2}}\\
    &\cdot(-n_{m,\boldsymbol{k}}+n_{n,\boldsymbol{k}+\boldsymbol{q}})].
\end{split}
\end{equation}
At zero temperature, $n_{m,\boldsymbol{k}}=0$ and only the first sum contributes.
In the main text, we utilize the low-energy, low-frequency scaling of $\mathrm{Re}\chi^{zz}(q,\omega)$ as $q^2/E_g$. This understanding is straightforward. The $b$-spinons form a nearly dispersionless, Landau-level-like spectrum in momentum space, where the low-energy fluctuations are primarily governed by the lowest energy level $E_g$. In other words, $E_g$ is the only relevant energy scale in this regime. We can also verify this numerically as shown in \figref{chibscaling}. The mean-field result is carried out for $t=2J$.
\section{Renormalization group analysis of the superconducting transition}
\label{RG}
In the superconducting phase, holons condense uniformly, allowing us to express $h_i$ as $\sqrt{n_i}e^{i\theta_i}$, where $\sqrt{n_i}$ and $\theta_i$ represent the amplitude and phase respectively. In the long-wavelength limit, we have
\begin{equation}
\begin{split}
L_h&=\sum_i n_i^h[\partial_\tau \theta(i)-iA^s_0(i)-iA^e_0(i)]\\
&-2 t_h n_i^h
\cos(\partial_\alpha\theta(i)-A_\alpha^s(i)-A_\alpha^e(i))\\
&+\mu(\sum_i h_i^\dagger h_i-N\delta)+\frac{u}{2}\sum_i n_i^2.
\end{split}
\end{equation}
where the onsite repulsion term $\frac{u}{2}\sum_i n_i^2$ is introduced in order to properly describe the hard-core nature of the holons. Using the Villian expansion,
\begin{equation}
e^{\beta \cos x}=
\text{const.}\sum_{N_\alpha\in\mathbf{Z}}\exp[-\frac{\beta}{2}(x-2\pi N_\alpha)^2],
\end{equation}
and expanding the holon density around its saddle point, $n_i^h=\bar{n}^h+\delta n_i^h$, the action can be expanded up to order $O(\delta n^2, \delta \theta^2)$. After integrating out the holon density fluctuation, we have
\begin{equation}
\begin{split}
&L_h+L_{\text{MCS}}\\
=&\sum_i\{\frac{1}{2u}[\partial_\tau \theta(i)-iA^s_0(i)-iA^e_0(i)]^2+\bar{n}_i^h(\partial_\tau \theta(i)-iA^s_0(i)\\
-&iA^e_0(i))+t_h \bar{n}_i^h(\partial_\alpha\theta(i)-A_\alpha^s(i)-eA_\alpha^s(i)+2\pi N_\alpha)\\
+&\frac{i}{\pi} A_0^s(i) \boldsymbol{B}^h(i)-\frac{i}{\pi}\boldsymbol{A}^s(i)\cdot (\boldsymbol{E}^h(i)\times\hat{z})\}.
\end{split}
\end{equation}
Performing a gauge transformation to eliminate the holon phase field, $A^s_\alpha\rightarrow A_\alpha^s-\partial_\alpha\theta$, $iA^s_0\rightarrow iA_0^s-\partial_\tau\theta$, and integrating out $\boldsymbol{A}^s$, we arrive at the effective Lagrangian describing the physics of the holon vortex coupled to the holon gauge field,
\begin{equation}
\begin{split}
L_{\text{eff}}&=\frac{u}{2\pi^2}\sum_i (B^h-\pi  \bar{n}_i^h)^2+\sum _i \frac{1}{4 \pi ^2 t_h \bar{n}_h}(E_{\alpha }^h)^2\\
&-\frac{i}{\pi}\sum_i \epsilon^{\mu\nu \lambda}A_{\mu}^e\partial_\nu A_{\lambda }^h-2i\sum_i A_\mu^h J^{\text{vor}}_\mu,
\end{split}
\end{equation}
where $J^{\text{vor}}_\mu$ captures the singular part of the phase $\theta$, $J^{\text{vor}}_0=\epsilon^{\alpha\beta}\partial_\alpha N_\beta$, $J^{\text{vor}}_\alpha=-\epsilon^{\alpha\beta}\partial_0 N_\beta$. The above derivation closely follows the standard particle-vortex duality transformation \cite{Kogut}. To figure out the interaction among the vortices, we start by integrating out $A_0^h$ which couples to the vortex density directly. Recall that $A^h_0$ couples to $b$-spinon spins as well, we need to combine $L_\text{eff}$ and $L_s$. The integration leads to
\begin{equation}
L_\text{vor}=-\frac{t_{h}\bar{n}^h\pi}{2}\sum_{i,j} Q_{i}\frac{\ln(r_i-r_j)}{a} Q_j,
\end{equation}where $Q_i=2J_0^{\text{vor}}(i)+\sum_{\sigma}\sigma  b^\dagger_{i\sigma}b_{i\sigma}$ is the effective vorticity, combining the $\pm\pi$ flux spinon and the $\mp 2\pi$ flux holon vortex. Since the vortex carries a singular self-energy, we restrict ourselves to consider the system with zero total vorticity. The new composite objects called spinon-vortices interact with each other through a logarithmic interaction similar to the 2-dimensional electron gas.
	
Since each spin-vortex contains a spinon excitation inside the core, we need to consider the excitation energy of the spinon $E_g/2$ as the self energy of the spinon-vortex, and only consider the vorticity $Q_i=\pm1$. The effective action of the system now reads
\begin{equation}
S_\text{eff}=\frac{E_g}{2k_B T}\sum_i|Q_i|-\pi\frac{K}{4}\sum_{i,j} Q_{i}\frac{\ln(r_i-r_j)}{a} Q_j,
\end{equation}
where $K=\frac{2t_h \bar{n}^h}{k_{B}T}$ is the reduced stiffness and the fugacity is $y=e^{-E_{g}/2k_{B}T}$.

Following the textbook level derivation of the Kosterlitz-Thouless transition, we can derive the corresponding RG equations of stiffness and fugacity,
\begin{align}
\frac{dK^{-1}}{dl}&=g^{2}\pi^{3}y^{2}+O(y^{4}),\\
\frac{dy}{dl}&=(2-\pi\frac{K}{4})y+O(y^{3}).
\end{align}
Here, $y$ is replaced by $gy$. $g=4$ represents the 4-fold degeneracy of each site on the von Neumann lattice, arising from time reversal and bipartite lattice symmetries \cite{Weng.Ma.2014}. The replacement of $K$ by $K/4$ contrary to the standard KT analysis reflects the $\pi$ vorticity of each spinon vortex instead of $2\pi$. The transition occurs at $(K^{*},y^{*})=(8/\pi,0)$. The critical trajectory separating flows
to zero and infinite $y$ corresponds to
\begin{equation}
    K_c^{-1}-\pi/8=-\pi^2 y_c+O(y_c^2).
\end{equation}
Since $k_BT_c/2t_h \bar{n}^h=k_BT_c/2t_h \delta\ll1$, we ignore $K_c$, and the RG equation leads to
\begin{equation}
\frac{E_{g}}{k_{B}T_{c}}=2\ln (8\pi)\approx 6.44.
\end{equation}
We briefly comment here that the value 5.76 reported in Ref. \cite{mei_spin-roton_2010} is obtained in a different way. There, the authors substitute the fixed-point value $K^*=8/\pi$ directly into the integral of Eq.(24) in Ref. \cite{mei_spin-roton_2010}, without simultaneously replacing other occurrences of $K$ and $y$ by their fixed-point values. This leads to a self-consistent equation for the renormalized stiffness $K_R$, derived to leading order in $K$ and $y$. While this procedure differs from the standard RG analysis, it successfully captures the correct scaling behavior of $T_c$ and $E_g$, in good agreement with experiments. Thus, the two methods yield quantitatively different values but share the same physical essence.

\bibliography{Bibliography.bib}

@article{Weng.Ma.2014,
year = {2014},
title = {{Low-temperature pseudogap phenomenon: precursor of high-$T_c$ superconductivity}},
author = {Ma, Yao and Ye, Peng and Weng, Zheng-Yu},
journal = {New J. Phys.},
issn = {1367-2630},
doi = {10.1088/1367-2630/16/8/083039},
pages = {083039},
number = {8},
volume = {16}
}

@article{Weng.Sheng.1996,
year = {1996},
title = {{Phase String Effect in a Doped Antiferromagnet}},
author = {Sheng, D. N. and Chen, Y. C. and Weng, Z. Y.},
journal = {Phys. Rev. Lett.},
issn = {0031-9007},
doi = {10.1103/physrevlett.77.5102},
pmid = {10062714},
pages = {5102--5105},
number = {25},
volume = {77},
month = {12}
}

@article{Zaanen.Wu.2008,
year = {2008},
title = {{Sign structure of the $t$-$J$ model}},
author = {Wu, K. and Weng, Z. Y. and Zaanen, J.},
journal = {Phys. Rev. B},
issn = {1098-0121},
doi = {10.1103/physrevb.77.155102},
pages = {155102},
number = {15},
volume = {77}
}

@article{Weng.Kou.2003yuc,
  title = {Topological Gauge Structure and Phase Diagram for Weakly Doped Antiferromagnets},
  author = {Kou, Su-Peng and Weng, Zheng-Yu},
  journal = {Phys. Rev. Lett.},
  volume = {90},
  issue = {15},
  pages = {157003},
  numpages = {4},
  year = {2003},
  month = {Apr},
  publisher = {American Physical Society},
  doi = {10.1103/PhysRevLett.90.157003},
  url = {https://link.aps.org/doi/10.1103/PhysRevLett.90.157003}
}

@article{Weng.Ye.2011, 
year = {2011}, 
title = {{Confinement-Deconfinement Interplay in Quantum Phases of Doped Mott Insulators}}, 
author = {Ye, Peng and Tian, Chu-Shun and Qi, Xiao-Liang and Weng, Zheng-Yu}, 
journal = {Phys. Rev. Lett.}, 
issn = {0031-9007}, 
doi = {10.1103/physrevlett.106.147002}, 
pmid = {21561214}, 
eprint = {1007.2507}, 
pages = {147002}, 
number = {14}, 
volume = {106}
}

@article{mei_spin-roton_2010,
	title = {Spin-roton excitations in the cuprate superconductors},
	volume = {81},
	issn = {1098-0121, 1550-235X},
	doi = {10.1103/PhysRevB.81.014507},
	number = {1},
	journal = {Physical Review B},
	author = {Mei, J. W. and Weng, Z. Y.},
	month = jan,
	year = {2010},
	pages = {014507}
}

@article{song_thermal_2024,
	title = {Thermal {Hall} effect and neutral spinons in a doped {Mott} insulator},
	volume = {6},
	issn = {2643-1564},
	doi = {10.1103/PhysRevResearch.6.023328},
	number = {2},
	journal = {Physical Review Research},
	author = {Song, Zhi-Jian and Zhang, Jia-Xin and Weng, Zheng-Yu},
	month = jun,
	year = {2024},
	pages = {023328}
}

@article{Kou,
  title = {Mutual Chern-Simons effective theory of doped antiferromagnets},
  author = {Kou, Su-Peng and Qi, Xiao-Liang and Weng, Zheng-Yu},
  journal = {Phys. Rev. B},
  volume = {71},
  issue = {23},
  pages = {235102},
  numpages = {12},
  year = {2005},
  month = {Jun},
  publisher = {American Physical Society},
  doi = {10.1103/PhysRevB.71.235102},
  url = {https://link.aps.org/doi/10.1103/PhysRevB.71.235102}
}

@article{Zaanen2016,
	title = {Superconducting electrons go missing},
	volume = {536},
	issn = {0028-0836, 1476-4687},
	url = {http://www.nature.com/articles/536282a},
	doi = {10.1038/536282a},
	number = {7616},
	urldate = {2022-12-24},
	journal = {Nature},
	author = {Zaanen, Jan},
	month = aug,
	year = {2016},
	pages = {282--283},
}

@article{Emery1995,
	title = {Importance of phase fluctuations in superconductors with small superfluid density},
	volume = {374},
	issn = {1476-4687},
	url = {https://doi.org/10.1038/374434a0},
	doi = {10.1038/374434a0},
	number = {6521},
	journal = {Nature},
	author = {Emery, V. J. and Kivelson, S. A.},
	month = mar,
	year = {1995},
	pages = {434--437},
}

@article{Bozovic2016,
  title    = "Dependence of the critical temperature in overdoped copper oxides
              on superfluid density",
  author   = "Bo{\v z}ovi{\'c}, I and He, X and Wu, J and Bollinger, A T",
  journal  = "Nature",
  volume   =  536,
  number   =  7616,
  pages    = "309--311",
  month    =  aug,
  year     =  2016
}

@article{Li2021,
	title    = "Superconductor-to-metal transition in overdoped cuprates",
	author   = "Li, Zi-Xiang and Kivelson, Steven A and Lee, Dung-Hai",
	journal  = "npj Quantum Materials",
	volume   =  6,
	number   =  1,
	pages    = "36",
	month    =  apr,
	year     =  2021
}

@article{Weng1997,
  title = {Phase string effect in the \textit{t}-\textit{J} model: General theory},
  author = {Weng, Z. Y. and Sheng, D. N. and Chen, Y.-C. and Ting, C. S.},
  journal = {Phys. Rev. B},
  volume = {55},
  issue = {6},
  pages = {3894--3906},
  numpages = {0},
  year = {1997},
  month = {Feb},
  publisher = {American Physical Society},
  doi = {10.1103/PhysRevB.55.3894},
  url = {https://link.aps.org/doi/10.1103/PhysRevB.55.3894}
}

@article{Chen2005,
  title = {Spin dynamics in a doped-Mott-insulator superconductor},
  author = {Chen, W. Q. and Weng, Z. Y.},
  journal = {Phys. Rev. B},
  volume = {71},
  issue = {13},
  pages = {134516},
  numpages = {15},
  year = {2005},
  month = {Apr},
  publisher = {American Physical Society},
  doi = {10.1103/PhysRevB.71.134516},
  url = {https://link.aps.org/doi/10.1103/PhysRevB.71.134516}
}

@article{Weng1999,
  title = {Mean-field description of the phase string effect in the \textit{t}-\textit{J} model},
  author = {Weng, Z. Y. and Sheng, D. N. and Ting, C. S.},
  journal = {Phys. Rev. B},
  volume = {59},
  issue = {13},
  pages = {8943--8955},
  numpages = {0},
  year = {1999},
  month = {Apr},
  publisher = {American Physical Society},
  doi = {10.1103/PhysRevB.59.8943},
  url = {https://link.aps.org/doi/10.1103/PhysRevB.59.8943}
}

@article {Anderson1987,
	author = {Anderson, P. W.},
	title = "{The Resonating Valence Bond State in $\textrm{La}_2\textrm{CuO}_4$ and Superconductivity}",
	volume = {235},
	number = {4793},
	pages = {1196--1198},
	year = {1987},
	doi = {10.1126/science.235.4793.1196},
	publisher = {American Association for the Advancement of Science},
	issn = {0036-8075},
	URL = {https://science.sciencemag.org/content/235/4793/1196},
	journal = {Science}
}

@article{Lee2006,
  title = {Doping a Mott insulator: Physics of high-temperature superconductivity},
  author = {Lee, Patrick A. and Nagaosa, Naoto and Wen, Xiao-Gang},
  journal = {Rev. Mod. Phys.},
  volume = {78},
  issue = {1},
  pages = {17--85},
  numpages = {0},
  year = {2006},
  month = {Jan},
  publisher = {American Physical Society},
  doi = {10.1103/RevModPhys.78.17},
  url = {https://link.aps.org/doi/10.1103/RevModPhys.78.17}
}

@article{Uemura1989,
  title = {Universal Correlations between $\textit{T}_c$ and $\frac{{n}_{s}}{{m}^{*}}$ (Carrier Density over Effective Mass) in High-$\textit{T}_c$ Cuprate Superconductors},
  author = {Uemura, Y. J. and Luke, G. M. and Sternlieb, B. J. and Brewer, J. H. and Carolan, J. F. and Hardy, W. N. and Kadono, R. and Kempton, J. R. and Kiefl, R. F. and Kreitzman, S. R. and Mulhern, P. and Riseman, T. M. and Williams, D. Ll. and Yang, B. X. and Uchida, S. and Takagi, H. and Gopalakrishnan, J. and Sleight, A. W. and Subramanian, M. A. and Chien, C. L. and Cieplak, M. Z. and Xiao, Gang and Lee, V. Y. and Statt, B. W. and Stronach, C. E. and Kossler, W. J. and Yu, X. H.},
  journal = {Phys. Rev. Lett.},
  volume = {62},
  issue = {19},
  pages = {2317--2320},
  numpages = {0},
  year = {1989},
  month = {May},
  publisher = {American Physical Society},
  doi = {10.1103/PhysRevLett.62.2317},
  url = {https://link.aps.org/doi/10.1103/PhysRevLett.62.2317}
}

@article{Uemura1992,
  title = {Basic Similarities among Cuprate, Bismuthate, Organic, Chevrel-Phase, and Heavy-Fermion Superconductors Shown by Penetration-Depth Measurements},
  author = {Uemura, Y. J. and Le, L. P. and Luke, G. M. and Sternlieb, B. J. and Wu, W. D. and Brewer, J. H. and Riseman, T. M. and Seaman, C. L. and Maple, M. B. and Ishikawa, M. and Hinks, D. G. and Jorgensen, J. D. and Saito, G. and Yamochi, H.},
  journal = {Phys. Rev. Lett.},
  volume = {68},
  issue = {17},
  pages = {2712--2712},
  numpages = {0},
  year = {1992},
  month = {Apr},
  publisher = {American Physical Society},
  doi = {10.1103/PhysRevLett.68.2712},
  url = {https://link.aps.org/doi/10.1103/PhysRevLett.68.2712}
}

@article{Uemura2003,
title = {Superfluid density of high-$\textit{T}_c$ cuprate systems: implication on condensation mechanisms, heterogeneity and phase diagram},
journal = {Solid State Communications},
volume = {126},
number = {1},
pages = {23-38},
year = {2003},
note = {Proceedings of the High-$\textit{T}_c$ Superconductivity Workshop},
issn = {0038-1098},
doi = {https://doi.org/10.1016/S0038-1098(02)00665-8},
url = {https://www.sciencedirect.com/science/article/pii/S0038109802006658},
author = {Y.J. Uemura}
}

@article{10.1038/s41586-021-04251-2, 
year = {2022}, 
title = {{Unconventional spectral signature of Tc in a pure d-wave superconductor}}, 
author = {Chen, Su-Di and Hashimoto, Makoto and He, Yu and Song, Dongjoon and He, Jun-Feng and Li, Ying-Fei and Ishida, Shigeyuki and Eisaki, Hiroshi and Zaanen, Jan and Devereaux, Thomas P. and Lee, Dung-Hai and Lu, Dong-Hui and Shen, Zhi-Xun}, 
journal = {Nature}, 
issn = {0028-0836}, 
doi = {10.1038/s41586-021-04251-2}, 
pmid = {35082417}, 
pages = {562--567}, 
number = {7894}, 
volume = {601}}

@article{qianghua_wang,
  title = {Anisotropic Scattering Caused by Apical Oxygen Vacancies in Thin Films of Overdoped High-Temperature Cuprate Superconductors},
  author = {Wang, Da and Xu, Jun-Qi and Zhang, Hai-Jun and Wang, Qiang-Hua},
  journal = {Phys. Rev. Lett.},
  volume = {128},
  issue = {13},
  pages = {137001},
  numpages = {6},
  year = {2022},
  month = {Mar},
  publisher = {American Physical Society},
  doi = {10.1103/PhysRevLett.128.137001},
  url = {https://link.aps.org/doi/10.1103/PhysRevLett.128.137001}
}

@article{Broun_disorder,
  title = {Disorder and superfluid density in overdoped cuprate superconductors},
  author = {Lee-Hone, N. R. and Dodge, J. S. and Broun, D. M.},
  journal = {Phys. Rev. B},
  volume = {96},
  issue = {2},
  pages = {024501},
  numpages = {8},
  year = {2017},
  month = {Jul},
  publisher = {American Physical Society},
  doi = {10.1103/PhysRevB.96.024501},
  url = {https://link.aps.org/doi/10.1103/PhysRevB.96.024501}
}

@Article{Tromp2023,
author={Tromp, Willem O.
and Benschop, Tjerk
and Ge, Jian-Feng
and Battisti, Irene
and Bastiaans, Koen M.
and Chatzopoulos, Damianos
and Vervloet, Amber H. M.
and Smit, Steef
and van Heumen, Erik
and Golden, Mark S.
and Huang, Yinkai
and Kondo, Takeshi
and Takeuchi, Tsunehiro
and Yin, Yi
and Hoffman, Jennifer E.
and Sulangi, Miguel Antonio
and Zaanen, Jan
and Allan, Milan P.},
title={Puddle formation and persistent gaps across the non-mean-field breakdown of superconductivity in overdoped $\text{(Pb,Bi)}_{2}\text{Sr}_{2}\text{Cu}\text{O}_{6+\delta}$},
journal={Nature Materials},
year={2023},
month={Jun},
day={01},
volume={22},
number={6},
pages={703-709},
issn={1476-4660},
doi={10.1038/s41563-023-01497-1},
url={https://doi.org/10.1038/s41563-023-01497-1}
}

@article{Patrick_normal,
  title = {Normal-state properties of the uniform resonating-valence-bond state},
  author = {Nagaosa, Naoto and Lee, Patrick A.},
  journal = {Phys. Rev. Lett.},
  volume = {64},
  issue = {20},
  pages = {2450--2453},
  numpages = {0},
  year = {1990},
  month = {May},
  publisher = {American Physical Society},
  doi = {10.1103/PhysRevLett.64.2450},
  url = {https://link.aps.org/doi/10.1103/PhysRevLett.64.2450}
}

@article{PhysRevB.98.054506,
  title = {Optical conductivity of overdoped cuprate superconductors: Application to $\text{La}_{2-x}\text{Sr}_{x}\text{CuO}_{4}$},
  author = {Lee-Hone, N. R. and Mishra, V. and Broun, D. M. and Hirschfeld, P. J.},
  journal = {Phys. Rev. B},
  volume = {98},
  issue = {5},
  pages = {054506},
  numpages = {10},
  year = {2018},
  month = {Aug},
  publisher = {American Physical Society},
  doi = {10.1103/PhysRevB.98.054506},
  url = {https://link.aps.org/doi/10.1103/PhysRevB.98.054506}
}

@Article{Ayres2021,
author={Ayres, J.
and Berben, M.
and {\v{C}}ulo, M.
and Hsu, Y.-T.
and van Heumen, E.
and Huang, Y.
and Zaanen, J.
and Kondo, T.
and Takeuchi, T.
and Cooper, J. R.
and Putzke, C.
and Friedemann, S.
and Carrington, A.
and Hussey, N. E.},
title={Incoherent transport across the strange-metal regime of overdoped cuprates},
journal={Nature},
year={2021},
month={Jul},
day={01},
volume={595},
number={7869},
pages={661-666},
issn={1476-4687},
doi={10.1038/s41586-021-03622-z},
url={https://doi.org/10.1038/s41586-021-03622-z}
}

@article{ResonancemodeOD,
  title = {Two Resonant Magnetic Modes in an Overdoped High ${T}_{c}$ Superconductor},
  author = {Pailh\`es, S. and Sidis, Y. and Bourges, P. and Ulrich, C. and Hinkov, V. and Regnault, L. P. and Ivanov, A. and Liang, B. and Lin, C. T. and Bernhard, C. and Keimer, B.},
  journal = {Phys. Rev. Lett.},
  volume = {91},
  issue = {23},
  pages = {237002},
  numpages = {4},
  year = {2003},
  month = {Dec},
  publisher = {American Physical Society},
  doi = {10.1103/PhysRevLett.91.237002},
  url = {https://link.aps.org/doi/10.1103/PhysRevLett.91.237002}
}

@article{Lipscombe,
  title = {Persistence of High-Frequency Spin Fluctuations in Overdoped Superconducting $\text{La}_{2-x}\text{Sr}_{x}\text{CuO}_{4}$ ($x=0.22$)},
  author = {Lipscombe, O. J. and Hayden, S. M. and Vignolle, B. and McMorrow, D. F. and Perring, T. G.},
  journal = {Phys. Rev. Lett.},
  volume = {99},
  issue = {6},
  pages = {067002},
  numpages = {4},
  year = {2007},
  month = {Aug},
  publisher = {American Physical Society},
  doi = {10.1103/PhysRevLett.99.067002},
  url = {https://link.aps.org/doi/10.1103/PhysRevLett.99.067002}
}

@article{Capogna2007,
  title = {Odd and even magnetic resonant modes in highly overdoped $\text{Bi}_{2}\text{Sr}_{2}\text{Ca}\text{Cu}_{2}\text{O}_{8+\delta}$},
  author = {Capogna, L. and Fauqu\'e, B. and Sidis, Y. and Ulrich, C. and Bourges, P. and Pailh\`es, S. and Ivanov, A. and Tallon, J. L. and Liang, B. and Lin, C. T. and Rykov, A. I. and Keimer, B.},
  journal = {Phys. Rev. B},
  volume = {75},
  issue = {6},
  pages = {060502},
  numpages = {4},
  year = {2007},
  month = {Feb},
  publisher = {American Physical Society},
  doi = {10.1103/PhysRevB.75.060502},
  url = {https://link.aps.org/doi/10.1103/PhysRevB.75.060502}
}

@article{Pailhes2006,
  title = {Doping Dependence of Bilayer Resonant Spin Excitations in $(\text{Y},\text{Ca})\text{Ba}_{2}\text{Cu}_{3}\text{O}_{6+x}$},
  author = {Pailh\`es, S. and Ulrich, C. and Fauqu\'e, B. and Hinkov, V. and Sidis, Y. and Ivanov, A. and Lin, C. T. and Keimer, B. and Bourges, P.},
  journal = {Phys. Rev. Lett.},
  volume = {96},
  issue = {25},
  pages = {257001},
  numpages = {4},
  year = {2006},
  month = {Jun},
  publisher = {American Physical Society},
  doi = {10.1103/PhysRevLett.96.257001},
  url = {https://link.aps.org/doi/10.1103/PhysRevLett.96.257001}
}

@article{Kyle2005,
author = {Kyle M. Shen  and F. Ronning  and D. H. Lu  and F. Baumberger  and N. J. C. Ingle  and W. S. Lee  and W. Meevasana  and Y. Kohsaka  and M. Azuma  and M. Takano  and H. Takagi  and Z.-X. Shen },
title = {Nodal Quasiparticles and Antinodal Charge Ordering in $\text{Ca}_{2-x}\text{Na}_{x}\text{CuO}_{2}\text{Cl}_{2}$},
journal = {Science},
volume = {307},
number = {5711},
pages = {901-904},
year = {2005},
doi = {10.1126/science.1103627},
URL = {https://www.science.org/doi/abs/10.1126/science.1103627}}

@article{Chatterjee,
author = {Utpal Chatterjee  and Dingfei Ai  and Junjing Zhao  and Stephan Rosenkranz  and Adam Kaminski  and Helene Raffy  and Zhizhong Li  and Kazuo Kadowaki  and Mohit Randeria  and Michael R. Norman  and J. C. Campuzano },
title = {Electronic phase diagram of high-temperature copper oxide superconductors},
journal = {Proceedings of the National Academy of Sciences},
volume = {108},
number = {23},
pages = {9346-9349},
year = {2011},
doi = {10.1073/pnas.1101008108},
URL = {https://www.pnas.org/doi/abs/10.1073/pnas.1101008108}}

@article{Hourglass,
  title = {Hourglass like spin excitation in a doped Mott insulator},
  author = {Zhang, Jia-Xin and Chen, Chuan and Zhang, Jian-Hao and Weng, Zheng-Yu},
  journal = {Phys. Rev. Res.},
  volume = {6},
  issue = {1},
  pages = {013109},
  numpages = {12},
  year = {2024},
  month = {Jan},
  publisher = {American Physical Society},
  doi = {10.1103/PhysRevResearch.6.013109},
  url = {https://link.aps.org/doi/10.1103/PhysRevResearch.6.013109}
}

@article{Kogut,
  title = {An introduction to lattice gauge theory and spin systems},
  author = {Kogut, John B.},
  journal = {Rev. Mod. Phys.},
  volume = {51},
  issue = {4},
  pages = {659--713},
  numpages = {0},
  year = {1979},
  month = {Oct},
  publisher = {American Physical Society},
  doi = {10.1103/RevModPhys.51.659},
  url = {https://link.aps.org/doi/10.1103/RevModPhys.51.659}
}

@article{Weng_2011,
doi = {10.1088/1367-2630/13/10/103039},
url = {https://dx.doi.org/10.1088/1367-2630/13/10/103039},
year = {2011},
month = {oct},
publisher = {IOP Publishing},
volume = {13},
number = {10},
pages = {103039},
author = {Weng, Zheng-Yu},
title = {Superconducting ground state of a doped Mott insulator},
journal = {New Journal of Physics}
}

@Inbook{Carlson2008,
author="Carlson, E. W.
and Emery, V. J.
and Kivelson, S. A.
and Orgad, D.",
editor="Bennemann, K. H.
and Ketterson, John B.",
title="Concepts in High Temperature Superconductivity",
bookTitle="Superconductivity: Conventional and Unconventional Superconductors",
year="2008",
publisher="Springer Berlin Heidelberg",
address="Berlin, Heidelberg",
pages="1225--1348",
isbn="978-3-540-73253-2",
doi="10.1007/978-3-540-73253-2_21",
url="https://doi.org/10.1007/978-3-540-73253-2_21"
}

@article{Keimer.Pailhes.2004, 
year = {2004}, 
title = {{Resonant Magnetic Excitations at High Energy in Superconducting $\mathrm{YBa}_2 \mathrm{Cu}_3 \mathrm{O}_{6.85}$}}, 
author = {Pailhes, S. and Sidis, Y. and Bourges, P. and Hinkov, V. and Ivanov, A. and Ulrich, C. and Regnault, L. P. and Keimer, B.}, 
journal = {Phys. Rev. Lett}, 
issn = {0031-9007}, 
doi = {10.1103/physrevlett.93.167001}, 
pmid = {15525020}, 
pages = {167001}, 
number = {16}, 
volume = {93}
}

@article{Dogan.Hayden.2004, 
year = {2004}, 
title = {{The structure of the high-energy spin excitations in a high-transition-temperature superconductor}}, 
author = {Hayden, S. M. and Mook, H. A. and Dai, Pengcheng and Perring, T. G. and Dogan, F.}, 
journal = {Nature}, 
issn = {0028-0836}, 
doi = {10.1038/nature02576}, 
pmid = {15175744}, 
pages = {531--534}, 
number = {6991}, 
volume = {429}
}

@article{Slaveboson_normal_state,
  title = {Gauge theory of the normal state of high-$\textit{T}_c$ superconductors},
  author = {Lee, Patrick A. and Nagaosa, Naoto},
  journal = {Phys. Rev. B},
  volume = {46},
  issue = {9},
  pages = {5621--5639},
  numpages = {0},
  year = {1992},
  month = {Sep},
  publisher = {American Physical Society},
  doi = {10.1103/PhysRevB.46.5621},
  url = {https://link.aps.org/doi/10.1103/PhysRevB.46.5621}
}

@article{chen2024non,
  title = {Electric transport in doped Mott insulators dictated by a non-Ioffe-Larkin composition rule and spinons},
  author = {Chen, Chuan and Zhang, Jia-Xin and Song, Zhi-Jian and Weng, Zheng-Yu},
  journal = {Phys. Rev. B},
  volume = {111},
  issue = {16},
  pages = {165138},
  numpages = {13},
  year = {2025},
  month = {Apr},
  publisher = {American Physical Society},
  doi = {10.1103/PhysRevB.111.165138},
  url = {https://link.aps.org/doi/10.1103/PhysRevB.111.165138}
}

@dataset{han_2025_17829990,
  author       = {Han, Zeyu and
                  Song, Zhi-Jian and
                  Zhang, Jia-Xin and
                  Weng, Zheng-Yu},
  title        = {Data files for "Intrinsic phase fluctuations and
                   superfluid density in doped Mott insulators"
                  },
  month        = dec,
  year         = 2025,
  publisher    = {Zenodo},
  doi          = {10.5281/zenodo.17829990},
  url          = {https://doi.org/10.5281/zenodo.17829990},
}

\end{document}